\documentclass[apjl]{emulateapj}
\usepackage{color}

\slugcomment{}

\shorttitle{A New Method for Extragalactic Distances}
\shortauthors{Yoshii et al.}

\begin{document}

\title{A NEW METHOD FOR MEASURING EXTRAGALACTIC DISTANCES}

\author
{Yuzuru Yoshii\altaffilmark{1},
 Yukiyasu Kobayashi\altaffilmark{2},
 Takeo Minezaki\altaffilmark{1},
 Shintaro Koshida\altaffilmark{3},
 and Bruce A. Peterson\altaffilmark{4}
}

\altaffiltext{1}{Institute of Astronomy, School of Science, University
of Tokyo, 2-21-1 Osawa, Mitaka, Tokyo 181-0015, Japan; PI of the MAGNUM project; yoshii@ioa.s.u-tokyo.ac.jp}
\altaffiltext{2}{National Astronomical Observatory, 2-21-1 Osawa, Mitaka, Tokyo 181-8588, Japan}
\altaffiltext{3}{Center of Astro Engineering and Department of Electrical Engineering,
Pontificia Univercsidad Catolica de Chile, Av. Vicuna Mackenna 4868, Chile}
\altaffiltext{4}{Mount Stromlo Observatory, Research School of Astronomy and Astrophysics, Australian National University, Weston Creek P.O., ACT 2611, Australia}

\begin{abstract}
We have pioneered a new method for the measurement of extragalactic distances.
This method uses the time-lag between variations in the short wavelength and
long wavelength light from an active galactic nucleus (AGN),  
based on a quantitative physical model of dust reverberation that
relates the time-lag to the absolute luminosity of the AGN.
We use the large homogeneous data set from intensive monitoring observations
in optical and near-infrared wavelength bands
with the dedicated 2-m MAGNUM telescope to obtain the distances to
17 AGNs in the redshift range $z=0.0024$ to $z=0.0353$. These distance
measurements are compared with distances measured using Cepheid variable stars,
and are used to infer that $H_0$ $=$ 73 $\pm$ 3 (random) km s$^{-1}$ Mpc$^{-1}$.
The systematic error in $H_0$ is examined, and the uncertainty in the size 
distribution of dust grains is the largest source of the systematic error, 
which is much reduced for a sample of AGNs for which their parameter 
values in the model of dust reverberation are individually measured.
This AGN time-lag method can be used beyond 30 Mpc,
the farthest distance reached by extragalactic Cepheids, and can be extended
to high-redshift quasi-stellar objects.
\end{abstract}



\section{INTRODUCTION}

\citet{Hubb29} discovered that the universe is expanding
by finding a correlation between a galaxy's recession velocity and
its distance. Since then, a reliable estimate of
the expansion rate of the universe at the current epoch has
been a central subject in observational cosmology. This expansion
rate is denoted by the Hubble constant, $H_0$, and it characterizes
the nature of the universe, such as the age of the universe, $t_0=1/H_0$,
the observable size of the universe,  $R_0=c/H_0$, and the critical mass
density of the universe, $\rho_{\rm  {crit,0}}=3H_0^{2}/(8\pi G)$,
where $c$ is the speed of light, and $G$ is the gravitational constant. 

A variety of empirical distance-ladder methods have been proposed,
which determine the distance to a galaxy through a series of steps,
with each step calibrating the next more distant step, such as those
with final steps based upon the period--luminosity relation for Cepheid
variable stars, the maximum luminosity of type Ia supernovae, etc.
The results from these empirical methods have almost converged to
a value of $H_0$ at around 73 km s$^{-1}$ Mpc$^{-1}$ \citep{Free01,Free10}.

On the other hand, a physical method has the advantage over
the empirical methods in that it could, in principle, determine $H_0$
through a reasonable model parameterization without resorting to
an empirical distance ladder. The Sunyaev--Zel'dovich effect \citep{Birk99}
and the use of gravitational lensing \citep{Blan86} have been proposed.
However, their results for $H_0$ have not converged to agree with those
from the empirical methods, because uncertainties associated with
parameters in these particular methods do not allow an accuracy comparable
to the empirical methods \citep{Free10}. Consequently, other physical
methods that enable the measurement of extragalactic distances with
higher accuracy are eagerly sought.

In this Letter, we propose a new physical method, using a model with simple physics,
for extragalactic distance determination. Our model is based on
the dust reverberation in active galactic nuclei (AGNs).
We demonstrate the effectiveness of this method
by determining $H_0$ with a value and accuracy comparable to that
obtained by the Hubble Key Project using Cepheid variable stars.

\begin{figure*}
\epsscale{0.81}
\plotone{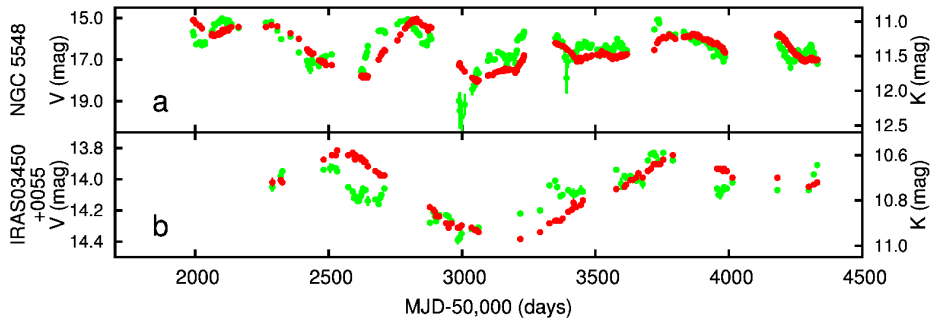}
\caption[The light curves of Seyfert galaxies]{
Light curves for two Seyfert 1 galaxies demonstrating the time-lag
between variations at short wavelength ($V$ band) and long wavelength ($K$ band).
Filled circles (colored red in the online version) represent
the $K$-band light curve, and open circles (filled green circles
in the online version) represent the $V$-band light curve.
Changes in the $V$-band are mimicked in the $K$-band
after a time-lag related to the $V$-band absolute magnitude.
The observations span more than 2000 days.
(a) Observations of NGC 5548 in the $V$ and $K$ bands.
(b) Observations of IRAS03450$+$0055 in the $V$ and $K$ bands.}\label{fig:lc}
\end{figure*}

\section{METHOD AND OBSERVATIONAL DATA}
An AGN consists
of a hot central engine surrounded by a cooler dust torus
\citep[e.g.,][]{Anto93,Urry95}. Dust near the central engine is sublimated,
creating a gap between the central engine and the inner radius of
the torus. Dust grains beyond the sublimation radius, $r_{\rm d}$, absorb
short wavelength light produced by the central engine, and re-emit
the absorbed energy at long wavelengths. Variations in the flux emitted
by the central engine are mimicked at long wavelengths by the dust
torus after a time delay, $\Delta t = r_{\rm d}/c$, corresponding to
the light travel time between the central engine and the inner radius
of the dust torus. Apparently, $r_{\rm d}$ is determined by the absolute luminosity,
$L$, of the central engine according to $r_{\rm d} \propto L^{1/2}$
\citep{Barv87,OH01,Mine04,Suga06}.
Thus, by measuring the time delay, $\Delta t$, we can obtain the absolute
luminosity of the AGN, and its distance.

We assume local energy balance
between the absorbed short wavelength ultraviolet (UV) radiation and
the re-emitted long wavelength near-infrared (NIR) radiation at the inner
radius of the dust torus:
\begin{equation}
\pi a^2 \int_{\rm UV}Q_{\nu}(a)\frac{L_{\nu }}{4\pi r^2_{\rm d}}d\nu=4\pi^2a^2\int_{\rm NIR}Q_{\nu}(a)B_{\nu}(T_{\rm d})d\nu,
\end{equation}
where $L_{\nu} = ({\nu}/{\nu}_{V})^{\alpha _{\rm UV}}L_{V}$
in the UV to optical region with the power-law index $\alpha _{\rm UV}$,
$L_{V}$ is the absolute luminosity in the $V$-band, $B_\nu(T_{\rm d})$ is
the Plank function for the characteristic dust sublimation temperature $T_{\rm d}$,
and $Q_\nu(a)$ is the absorption coefficient per dust grain of mean radius $a$
as a function of $\nu$ taken from literature \citep{Drai84,LD93}.
Values of the three parameters, $\alpha _{\rm UV}$ , $T_{\rm d}$, and $a$, are
set using observations of AGN. We have $\alpha_{\rm UV} = -0.5 \pm 0.2$
taken from a composite QSO spectrum \citep{VanB01,Davi07}.  
We adopt a rather large uncertainty of $\pm 0.2$ which well covers the 
observed range of QSO to QSO spectral index variation. With this $\alpha_{\rm UV}$, 
we truncate the UV spectrum at $0.03$ $\mu$m, considering the UV turnover of 
QSO spectrum \citep[e.g.,][]{Telf02,Kraw13} and the decrease of dust absorption efficiency at shorter 
wavelength below this truncation \citep{Drai84,LD93}. 
We have $T_{\rm d} = 1700 \pm 50$ K evaluated from the $H-K$, $J-K$, and $J-H$ color
temperatures of the variable NIR component for a sample of Seyfert 1 galaxies
observed with the MAGNUM telescope \citep{Tomi05,Tomi06}
\footnote{
Since the time delay is measured using the NIR to optical flux variations, the dust 
temperature near $r_{\rm d}$ is evaluated from the NIR color temperatures 
of the variable component rather than the average or single-epoch NIR spectral 
energy distributions of the whole component which contains the contribution from 
the dust of lower temperatures at larger radii beyond $r_{\rm d}$ \citep[e.g.,][]{Kish07,Land11}.
AGNs in our sample show no burst-like flux variation but 
repeat the brighter and fainter states in turn (Figure 1), keeping their 
NIR colors almost constant \citep{Tomi05,Suga06}. Therefore, the dust temperature 
of the variable component is regarded as staying almost constant during the flux variation. 
Note that the color 
temperatures can be determined without subtracting the host-galaxy flux, 
which would often be an origin of large uncertainty.
}.
The value for $T_{\rm d}$ is consistent with the condensation temperature
for solid carbon \citep{Huff97,Salp77}.
Since dust grains other than graphite grains
sublimate below this temperature, the absorption coefficient of graphite grains is 
adopted for $Q_\nu(a)$.
The distribution of $a$ is assumed to have a power-law form of $f(a) = Ka^{-p}$ 
with $p=2.75$, $a_{\rm min}=0.005\ \mu$m, and $a_{\rm max}=0.20\ \mu$m (cf. \S3).
The mean grain size for this distribution
is larger than that in the interstellar medium,
based on the analysis of UV extinction curves
of radio-quiet AGNs \citep{Gask04}.

\begin{deluxetable*}{lcccccccc}
\tablewidth{0pt}
\tabletypesize{\scriptsize}
\tablecaption{List of Target Active Galactic Nuclei}
\tablehead{
\colhead{Object}     & \colhead{R. A.} & \colhead{Decl.} &
\colhead{$z$\tablenotemark{a}} & \colhead{$v$\tablenotemark{b}} &
\colhead{$m_V$\tablenotemark{c}} & \colhead{$A_V$\tablenotemark{d}} &
\colhead{$\Delta t$\tablenotemark{e}} & \colhead{$d$\tablenotemark{f}} \\
                  &            &             &        &
\colhead{(km s$^{-1}$)} & \colhead{(mag)} & \colhead{(mag)} & \colhead{(days)} & \colhead{(Mpc)}}
\startdata
Mrk 335           & 00 06 19.5 & $+$20 12 10.5 & 0.0258 & ~\,8996.07$\pm $1383.98      & 14.59$\pm $0.02 & 0.118 & 139.2$\pm $16.4     & 145.6 $\pm $ 17.3 \\
Mrk 590           & 02 14 33.6 & $-$00 46 00.1 & 0.0264 & ~\,7176.93$\pm $~\,551.52    & 16.56$\pm $0.08 & 0.124 & ~\,36.8$\pm $~\,2.7 & ~\,95.2 $\pm $ ~\,7.0 \\
IRAS 03450$+$0055 & 03 47 40.2 & $+$01 05 14.0 & 0.0310 & ~\,7787.03$\pm $1347.99      & 14.74$\pm $0.02 & 0.660 & 103.3$\pm $~\,4.7   & ~\,90.4 $\pm $ ~\,4.1 \\
Akn 120           & 05 16 11.4 & $-$00 08 59.4 & 0.0327 & ~\,8041.23$\pm $1695.01      & 13.82$\pm $0.02 & 0.426 & 135.4$\pm $16.8     & ~\,86.5 $\pm $ 10.8 \\
MCG $-$08$-$11$-$011 & 05 54 53.5 & $+$46 26 22.0 & 0.0205 & ~\,6541.04$\pm $~\,331.25 & 15.11$\pm $0.06 & 0.720 & ~\,92.6$\pm $13.7   & ~\,93.1 $\pm $ 13.9 \\
Mrk 79            & 07 42 32.8 & $+$49 48 34.8 & 0.0222 & ~\,6753.90$\pm $~\,~\,75.14  & 15.11$\pm $0.04 & 0.235 & ~\,71.9$\pm $~\,3.3 & ~\,90.5 $\pm $ ~\,4.2 \\
Mrk 110           & 09 25 12.9 & $+$52 17 10.5 & 0.0353 & 10240.19$\pm $~\,587.93      & 15.28$\pm $0.04 & 0.043 & ~\,88.5$\pm $~\,6.0 & 132.5 $\pm $ ~\,9.1 \\
NGC 3227          & 10 23 30.6 & $+$19 51 54.0 & 0.0039 & ~\,1214.99$\pm $~\,~\,35.56  & 14.56$\pm $0.06 & 0.075 & ~\,13.9$\pm $~\,0.6 & ~\,14.5 $\pm $ ~\,0.6 \\
NGC 3516          & 11 06 47.5 & $+$72 34 07.0 & 0.0088 & ~\,3836.38$\pm $~\,844.97    & 15.09$\pm $0.09 & 0.140 & ~\,51.2$\pm $~\,9.6 & ~\,66.1 $\pm $ 12.4 \\
Mrk 744           & 11 39 42.5 & $+$31 54 33.0 & 0.0089 & ~\,4171.02$\pm $1031.14      & 17.23$\pm $0.06 & 0.079 & ~\,20.9$\pm $~\,2.2 & ~\,74.3 $\pm $ ~\,7.8 \\
NGC 4051          & 12 03 09.6 & $+$44 31 52.8 & 0.0023 & ~\,~901.24$\pm $~\,128.74    & 14.85$\pm $0.09 & 0.043 & ~\,14.6$\pm $~\,0.5 & ~\,17.6 $\pm $ ~\,0.6 \\
NGC 4151          & 12 10 32.6 & $+$39 24 20.6 & 0.0033 & ~\,1409.56$\pm $~\,299.65    & 13.45$\pm $0.09 & 0.092 & ~\,47.2$\pm $~\,0.7 & ~\,29.2 $\pm $ ~\,0.4 \\
NGC 4593          & 12 39 39.4 & $-$05 20 39.4 & 0.0090 & ~\,3540.76$\pm $~\,545.81    & 15.23$\pm $0.04 & 0.082 & ~\,43.6$\pm $~\,1.9 & ~\,61.9 $\pm $ ~\,2.7 \\
NGC 5548          & 14 17 59.5 & $+$25 08 12.4 & 0.0172 & ~\,6115.16$\pm $~\,486.34    & 15.81$\pm $0.03 & 0.068 & ~\,49.5$\pm $~\,0.8 & ~\,92.5 $\pm $ ~\,1.5 \\
Mrk 817           & 14 36 22.1 & $+$58 47 39.4 & 0.0315 & ~\,8779.59$\pm $1023.15      & 14.83$\pm $0.02 & 0.022 & ~\,87.9$\pm $~\,8.1 & 107.6 $\pm $ 10.0 \\
Mrk 509           & 20 44 09.7 & $-$10 43 24.5 & 0.0344 & ~\,9026.24$\pm $1419.98      & 13.87$\pm $0.03 & 0.190 & 144.5$\pm $~\,9.1   & 105.4 $\pm $ ~\,6.7 \\
NGC 7469          & 23 03 15.6 & $+$08 52 26.4 & 0.0163 & ~\,3967.24$\pm $~\,906.33    & 14.58$\pm $0.06 & 0.228 & ~\,48.4$\pm $~\,1.5 & ~\,47.7 $\pm $ ~\,1.5 \\
\enddata
\label{tabdata}
\tablenotetext{a}{The heliocentric redshift from the NASA/IPAC Extragalactic Database (NED).}
\tablenotetext{b}{The the cosmic recession velocity: $cz$ corrected for galactic rotation and
the local velocity flow. The $\pm$ figures represent the uncertainty in the models
of the local velocity flow.}
\tablenotetext{c}{The mean of the $V$-band fluxes generated at equal intervals of time, after
excluding the data which statistically show no flux variation in the time bins,
with the $1\sigma$ error.}
\tablenotetext{d}{The galactic extinction correction for $m_V$.}
\tablenotetext{e}{The time-lag with $1\sigma$ errors as described in the text.}
\tablenotetext{f}{The distance, $d$, as determined from Equation (2).}
\end{deluxetable*}

In order to obtain $\Delta t_{\rm AGN}$ in the rest frame of the AGN,
two cosmological effects must be considered. The observed $\Delta t$
must be corrected for time dilation, a factor of $(1+z)^{-1}$, where $z$
is the redshift. We also consider an effect that arises from the temperature
gradient in the dust torus and the shift of the $K$-band, where our
observations are made, to a different wavelength in the reference frame
of the AGN according to $\lambda_{\rm AGN} = \lambda_{K}/(1 + z)$. At higher redshifts,
we observe shorter wavelengths emitted by the AGN from a hotter part of
the dust torus that is closer to the central engine. This effect makes
the observed $\Delta t$ shorter than would be the case if the same AGN
was observed at a smaller redshift. We formulate the correction to our
$K$-band observations empirically, by using the mean delay of $H$-band
to $K$-band, which is about 0.3 times the delay of $V$-band to $K$-band,
as obtained from MAGNUM telescope observations of nearby Seyfert 1
galaxies \citep{Tomi06}. Interpolating this band-dependent delay linearly
with redshift, and correcting for the time dilation, we obtain
the correction formula $\Delta t_{\rm AGN} = \Delta t (1 + 1.24z)/(1 + z)$.
For the redshift range in this Letter, $z \le 0.035$, the combined
corrections are less than 1\% and are unimportant.

The MAGNUM project was initiated in order to measure $\Delta t$ for AGNs
over a range of redshifts \citep{Yosh02,YKM03}. The observations were
made with a multicolor imaging photometer \citep{Koba98} mounted on
a dedicated 2-m telescope located at the summit of Haleakala
on the Hawaiian Island of Maui. Monitoring observations from 2001
through 2007 were made simultaneously in two wavelength bands,
the optical $V$ (0.55 $\mu$m) band and the NIR $K$ (2.2 $\mu$m) band,
to determine $\Delta t$.
Variations in the short wavelength flux which heats the dust grains
is synchronous with variations in the $V$-band \citep[e.g.,][]{Wink97,Saka10}
and dust near the sublimation temperature emits the most energy in
the $K$-band \citep[e.g.,][]{Barv90,Koba93}. In Figure 1, we present
light curves in the $V$-band and $K$-band for NGC 5548 and
IRAS 03450$+$0055, as examples.
The flux variations in the $V$-band and $K$-band resemble each other
in shape, with the $K$-band flux variations delayed with respect to
those in the $V$-band.
In this Letter, we use the time-delay data for 17 Seyfert 1 galaxies
obtained by the MAGNUM project. 
We only briefly describe the procedures of our data
reduction and analysis here, because the details will be
presented elsewhere (S. Koshida et al., in preparation).

The time delay, $\Delta t$, was obtained by cross-correlating the $V$-band
and $K$-band light curves using a bi-directional interpolation scheme
which weights the observed flux data more than a simulated interpolation
scheme with respect to time \citep{Suga06}.

Before taking the cross correlation between the $V$-band and
$K$-band light curves, we subtracted the accretion disk component
from the observed $K$-band flux to derive the dust torus component
in the $K$ band \citep{Kosh09}.
Emission from the central accretion disk extends into
the NIR wavelengths \citep{Mine06,Tomi06,Kish08}.
Variations of this component appear in the $K$ band, and
make a measured $\Delta t$ shorter than the light travel time.
We correct by estimating the contaminating flux as
$f_{K} = (\nu_{V}/\nu_{K})^{\alpha_{VK}} f_{V}$
from the simultaneous $V$-band measurement, and subtracting it
from the $K$-band measurement.
We use $\alpha_{VK} = 0.1 \pm 0.11$ obtained from a cross
multiple regression analysis of our sample of nearby
Seyfert 1 galaxies \citep{Tomi05,Tomi06}.

\begin{figure}
\epsscale{1.15}
\plotone{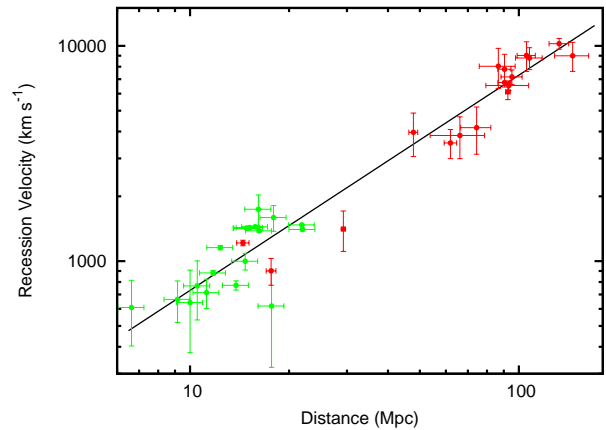}
\caption[The Hubble diagram]{
Velocity vs. distance for galaxies with distances measured using
Cepheid variable stars and for galaxies with distances measured
using the AGN time-lag method in this Letter. 
Open circles (filled green circles in the online version) represent
the data based on Cepheid variable stars as part of the Hubble Space Telescope
Key Project \citep{Free01},
and filled circles (colored red in the online version) represent our results.
All recession velocities have been derived from redshift
measurements that have been corrected for the local velocity
field using the same procedures.
The inclined line represents the relation $v=H_0 d$ where
$H_0 = 73$ km s$^{-1}$ Mpc$^{-1}$. The horizontal error bars represent
$1\sigma$ errors in the distance. The vertical error bars represent
the uncertainty in the local velocity field correction.}
\end{figure}

In the light curves of 17 Seyfert 1 galaxies, we found 49 single incidents
of a flux maximum or minimum which could be used to measure $\Delta t$.
We give the mean value of $\Delta t$ for those galaxies which underwent
several incidents. In Table 1 we give, for each of the 17 Seyfert 1 galaxies
that we monitored, our mean observed apparent magnitude, $m_V$,
the time delay, $\Delta t$ in days, and the luminosity distance, $d$ in Mpc,
from Equation (2)
\begin{equation}
d=\Delta t_{\rm AGN}\times 10^{0.2(m_V-A_V-k_V-25+g)},
\label{eqn_dl}
\end{equation}
where $A_V$ is Galactic extinction \citep{Schl98}, $k_V$ is the $K$-correction,
and $g = 10.60$. The calibration factor, $g$, was calculated from Equation (1)
using the physical model with the parameters described above. The $K$-correction
was calculated assuming a power law spectrum in the UV to optical range,
$f_{\nu} \propto \nu^{\alpha_{\rm UV}}$, to obtain
$k_V = -2.5(1 + \alpha_{\rm UV})\log(1 + z)$, where $\alpha_{\rm UV} = -0.5$.

The heliocentric redshift, $z$, is taken from published \ion{H}{1} measurements,
the cosmic recession velocity, $v$, is the heliocentric velocity, $cz$,
corrected for galactic rotation and the local velocity flow using
the average of two different models of the local velocity flow \citep{Moul00,Tonr00},
as was done for the Cepheid distance determinations \citep{Free01}.
The $\pm$ figures given for $v$ in Table 1 represent the uncertainty
in the flow models and are half the difference between the correction
to $v$ obtained from the two models.

\section{RESULTS AND DISCUSSION}
In Figure 2, we present the velocity-distance diagram to compare the extragalactic
distances measured using our AGN time-lag method with distances obtained from
Cepheid variable stars. Our AGN time-lag method determines the distance, $d$,
through a reasonable physical parameterization, and directly provides
the first step in the distance ladder. We obtain the Hubble constant,
$H_0 = 73 \pm 3$ km s$^{-1}$ Mpc$^{-1}$ from a least squares fit to 
$v = H_0d$ using our AGN time-lag distances. The Hubble constant found from
empirically calibrated Cepheid distances is $H_0 = 75 \pm 10$ \citep{Free01}.
Our AGN time-lag method, which is based upon a physical model with no empirical
calibration, agrees well with the Cepheid distances and other empirical
distance ladders \citep{Free01,Free10}, but extends to galaxies 10 times more
distant than where Cepheid distances can be measured, to where the cosmic
recession velocity is not so badly afflicted by the local velocity flow.

The systematic error in $H_0$ is estimated by changing the parameter values 
of $\alpha_{\rm UV}$ and $T_{\rm d}$ in their respective ranges of uncertainty, 
and by changing the distribution of $a$, assumed to have a power-law form of 
$f(a) = Ka^{-p}$.
\footnote{
The observer's viewing angle has also been considered as a possible 
source of the systematic error \citep[e.g.,][]{KW11}. However, 
we do not explicitly consider this possibility here, because we have not 
found any systematic difference in the time lag measurements for our target 
AGNs of different Seyfert subclasses (S. Koshida et al., in preparation).  
}
Our calculation gives 
$\Delta H_{0,\alpha}=\pm 5$ km s$^{-1}$ Mpc$^{-1}$ and 
$\Delta H_{0,T_{\rm d}}=\pm 3$ km s$^{-1}$ Mpc$^{-1}$ for the uncertainty 
in $\alpha_{\rm UV}$ and $T_{\rm d}$, respectively.
The systematic error in $H_0$ from the uncertainty in $f(a)$ is 
estimated using two extreme size distributions, such as
the standard ``MRN'' distribution in the local interstellar medium for our Galaxy 
\citep[$p=3.5$, $a_{\rm min}=0.005\ \mu$m, $a_{\rm max}=0.25\ \mu$m;][]{Math97}, and the larger grain enhanced distribution for AGNs 
($p=2.05$, $a_{\rm min}=0.005\ \mu$m, $a_{\rm max}=0.20\ \mu$m), because
smaller dust grains are more efficiently sublimated by UV radiation from 
the central engine \citep{Gask04}. We adopt an intermediate value of 
$p=2.75$ with $a_{\rm min}=0.005\ \mu$m and $a_{\rm max}=0.20\ \mu$m
as our standard grain size distribution.
NIR flux-weighted averaging scheme over the full range of $a$ is then
applied to Equation (1), and the calibration factor $g$ is calculated
for the two extreme distributions as well as 
the intermediate one. The uncertainty in $g$ is within 
the limits of $\Delta g=\pm 0.5$, so that $\Delta H_{0,a}$ is at most 
$\pm 0.1$ dex from Equation (2), which is currently the largest source 
of the systematic error in $H_0$.  
This error in $H_0$ is much reduced for a sample of AGNs for which 
their $\alpha_{\rm UV}$, $T_{\rm d}$, and $f(a)$ are individually 
measured, because the target to target variation of these parameters
also contributes to the random error in the $H_0$ fitting. 
In particular, if monitoring observations of high-redshift AGNs were 
obtained, their UV--optical spectrum could be determined directly from 
ground based spectroscopic observations, and the accuracy of the $H_0$ 
determination would be significantly improved. Alternatively,
if many of target AGNs were calibrated by other reliable
distance indicators such as Cepheids and type Ia supernovae, their 
calibrated distances would not only provide a cross check on our method 
of distance determination, but also independently constrain the parameter 
values in a physical model of dust reverberation.

Two other reverberation methods of distance determination
of AGN have been proposed.
One is a method based on the reverberation mapping of
the broad line emitting region (BLR) in AGN \citep{Wats11,Cze03},
using an empirical relation between the radial distance of BLR
from the AGN center and the AGN luminosity.
Although the radius-luminosity relation of BLR is well established
now \citep{Bent13}, the exact size of BLR cannot be
predicted theoretically.
Therefore, this method remains as an empirical method in the sense
that it must use empirical distance-ladder methods for calibration. 

Another is a method based on the wavelength-dependent time delay of UV--optical
continuum emission from the central accretion disk in AGN \citep{Coll99,Cack07}.
This method is a physical one, like our method, which could, in principle,
determine the distance without any distance calibration. However, this method
gives $H_0=$42--44 km s$^{-1}$ Mpc$^{-1}$, which is about a factor of
1.7 smaller than current standard estimates, and even below the recent
lower estimate of 67 km s$^{-1}$ Mpc$^{-1}$ based on Planck measurements of
the cosmic microwave background temperature \citep{Plan13}.
Thus, this method seems to have some difficulties to be
resolved, such as the adequacy of modeling the X-ray
reprocessing for the flux variation of UV--optical continuum
emission. In fact, while some AGNs show a delay of the
optical flux variation behind that of X-ray, which is consistent
with the X-ray reprocessing, some others show a delay
of the X-ray variation behind that of optical, or they show
a poor correlation between the X-ray and optical variations
\citep[e.g.,][]{Uttl06}.

In summary,
we have demonstrated that our AGN time-lag method can be used to measure
extragalactic distances beyond what is possible with Cepheids, and we have 
obtained a value of the Hubble constant with our AGN time-lag method of
$H_0 = 73 \pm 3\ ({\rm random})$ km s$^{-1}$ Mpc$^{-1}$ for a sample of 
17 Seyfert 1 galaxies observed with the MAGNUM telescope. We suggest that 
this method can be used with QSOs to study the dark energy in the universe 
beyond what is possible with type Ia supernovae.

\acknowledgments

We thank the staff at the Haleakala Observatories for their help with
facility maintenance. This research has been supported partly by
the Grants-in-Aid of Scientific Research (10041110, 10304014, 11740120,
12640233, 14047206, 14253001, 14540223, 16740106, and 25287062) and the COE Research
(07CE2002) of the Ministry of Education, Science, Culture and Sports of Japan.


\end{document}